\documentclass[conference]{IEEEtran}
\IEEEoverridecommandlockouts
\usepackage{lipsum}
\usepackage{amsmath,amssymb,amsfonts}
\usepackage{algorithmic}
\usepackage{graphicx}
\usepackage{subcaption}
\usepackage{textcomp}
\usepackage{xcolor}
\usepackage{graphicx}
\usepackage[english]{babel}
\usepackage[noabbrev]{cleveref}
\usepackage{biblatex} 
\def\BibTeX{{\rm B\kern-.05em{\sc i\kern-.025em b}\kern-.08em
    T\kern-.1667em\lower.7ex\hbox{E}\kern-.125emX}}
\begin{document}

\title{Change Management using Generative Modeling on Digital Twins}
\author{\IEEEauthorblockN{Nilanjana Das}
\IEEEauthorblockA{\textit{C.S.E.E. Dept.} \\
\textit{University of Maryland,}\\
\textit{Baltimore County}\\
ndas2@umbc.edu}
\and
\IEEEauthorblockN{Anantaa Kotal}
\IEEEauthorblockA{\textit{C.S.E.E. Dept.} \\
\textit{University of Maryland,}\\
\textit{Baltimore County}\\
anantak1@umbc.edu}
\and
\IEEEauthorblockN{Daniel Roseberry}
\IEEEauthorblockA{\textit{C.S.E.E. Dept.} \\
\textit{University of Maryland,}\\
\textit{Baltimore County}\\
danielr2@umbc.edu}
\and
\IEEEauthorblockN{Anupam Joshi}
\IEEEauthorblockA{\textit{C.S.E.E. Dept.} \\
\textit{University of Maryland,}\\
\textit{Baltimore County}\\
joshi@cs.umbc.edu}
}
\maketitle

\begin{abstract}
A key challenge faced by small and medium-sized business entities is securely managing software updates and changes. Specifically, with rapidly evolving cybersecurity threats, changes/updates/patches to software systems are necessary to stay ahead of emerging threats and are often mandated by regulators or statutory authorities to counter these. However, security patches/updates require stress testing before they can be released in the production system. Stress testing in production environments is risky and poses security threats. Large businesses usually have a non-production environment where such changes can be made and tested before being released into production. Smaller businesses do not have such facilities. In this work, we show how ``digital twins", especially for a mix of IT and IoT environments, can be created on the cloud. These digital twins act as a non-production environment where changes can be applied, and the system can be securely tested before patch release. Additionally, the non-production digital twin can be used to collect system data and run stress tests on the environment, both manually and automatically. In this paper, we show how using a small sample of real data/interactions, Generative Artificial Intelligence (AI) models can be used to generate testing scenarios to check for points of failure.
\end{abstract}


\section{Introduction}
A major challenge in cybersecurity today is staying ahead of emerging threats and addressing old vulnerabilities before they are exploited. Constant updates/patches are critical in the maintenance phase of a product development life cycle. They are also some of the most effective ways to counter security threats that exploit existing vulnerabilities. Some of these updates have been made mandatory by the Government. In an October 2022 report on Cybersecurity Awareness \cite{Cybersec0:online}, NIST reiterates why constant updates are our best defense against cyber threats. However, releasing a security patch for a product or system is risky. Without proper testing, a security patch can break the existing infrastructure or create new vulnerabilities. To provide seamless integration with the production environment, product developers and system designers typically run these patches/updates through a series of testing in a pre-production environment that simulates the production environment. Larger business entities have the resources to build both production and non-production environments that can be used for pre-release testing and system integration. However, maintaining an independent resource for non-production testing can be expensive. Smaller business entities have limited resources and often do not have a separate non-production environment. To solve this problem, a virtual copy, a ``digital twin", can be created of the existing system \cite{bhatt2022digital, lin2021efficient, incki2012survey}. The twin behaves as a non-production environment for testing patches prior to production release.  In this paper, we focus on both single machines and IoT based Cyberphysical systems. In developing a digital twin for an IT and IoT device, the process needs a platform to support the cloud environment and an agent to facilitate the capture and upload of information to the cloud platform.

After the digital twin is created, manual tests can be conducted on this new cloud-based system. However, running manual tests can be time-consuming. One approach is to capture the user interaction/input to the system, and then replay it on the twin. However, before releasing a patch/update into the production environment, the system behavior has to be observed over long periods of time with different inputs after application of the patch/update to determine it as safe for production. Inputs to the system captured during normal use in the production environment before the update will form the core of the tests. However, in any limited capture, we run the risk that only a subset of the possible inputs have been captured. Our key innovation is using AI to overcome this limitation. Input data captured as described above can be used to train generative machine learning (ML) models  to generate new, synthetic input data. The synthetic system data can be used to create a wide array of testing scenarios and reduce the need for manual test design.

In this paper, we propose a framework to capture a system image of the IT and IoT devices, upload them to a cloud and create a ``digital twin" that acts like a non-production environment. Furthermore, to enhance testing in the digital twin, we record data in different formats from IT and IoT devices and use the data to train ML models and generate a multitude of testing scenarios in a digital twin. We demonstrate our framework by creating digital twins of IT and IoT devices on the cloud and using them for patch/update testing.

\section{Background and Related Work}
\label{sec_Background}
We use digital twins to facilitate smaller business entities with a non-production environment. The digital twin provides a platform to the end-users such that they can perform automated or manual tests on it \cite{bhatt2022digital, lin2021efficient, incki2012survey}. These tests will help to understand the functioning of the system after a patch/change is applied to it. Jagatheesaperumal et al. \cite{10017413} said that a digital twin enables international firms to create digital replicas of their operations, assets and products to improve performance and upkeep. Barricelli et al. \cite{8901113} and Tao et al. \cite{tao2018digital} focuses on digital twins in an industry such as their applications or how much they have been developed in this sector. Our work extends the use of digital twins for IT and IoT devices with the most common Operating Systems, like Windows or Ubuntu. We build twins that can help evaluate security implications of change/patch management, and use it to stage a pre-production testing environment where automated or manual tests can be executed.Furthermore, we use the data collected in the system to automatically create additional testing scenarios using Generative AI for synthetic system data.

Software updates require testing in the organization's IT infrastructure, networks, and systems, before they can be deployed to production. These requires constant monitoring of the system health in the pre-production environment. Modern tools integrate Machine Learning (ML) models to enhance our system monitoring capabilities \cite{da2019internet, ucci2019survey, ding2018survey, piplai2020using, dasgupta2020comparative}. However, vigorous testing, specifically with the use of ML techniques, require significant amount of data which is infeasible in limited testing environments. The data collected over digital twin can be used to generate synthetic system data serves several purposes. It can be used to create functional tests to test the system is working fine after patch is applied. In general, there is a lot of evidence of GANs being used for synthetic data generation and translation in image and text data \cite{brock2018large, isola2017image, zhang2017stackgan, wang2018high}. However, the properties of system or device data makes it distinct from image and text data. A conditional generator model can address the issue of  mixed attributes in tabular data by seeking to minimize the distance between generated and real data given a fixed value of the discrete variable \cite{arjovsky2017wasserstein, xu2018synthesizing, xu2019modeling, kotal2022privetab}. System data, specifically network activity data, generated using conditional GANs have been shown to replace actual datasets in downstream tasks like network intrusion detection \cite{shahid2020generative, 8936224}.

\begin{figure*}
        \centering
        \includegraphics[scale=0.7]{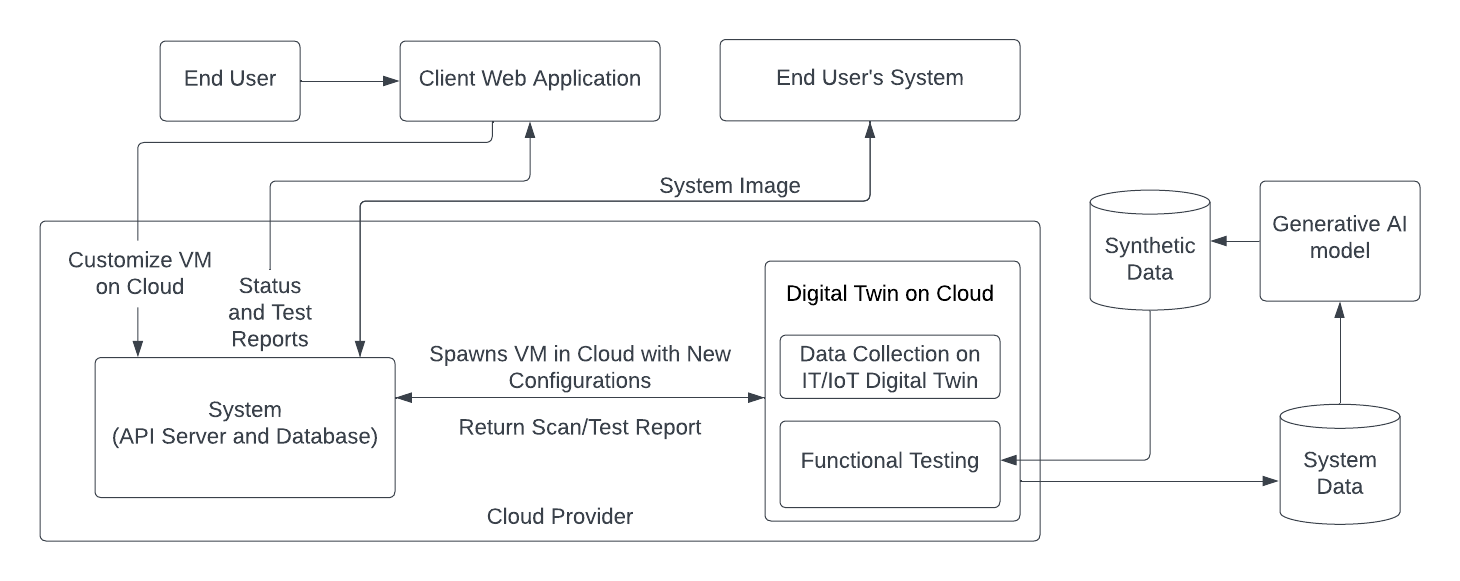}
    \caption{System Architecture for creating and testing on Digital Twins of IT and IoT devices}
      \label{Fig_System_Architecture}

\end{figure*}

\section{Architecture and System Design}
\label{sec_arch}
The digital twin is a virtualized copy of the existing system that can be hosted on the cloud. The patch/update can be applied on the digital twin. The system logs of the digital twin can be observed for any changes in system behavior as a result of the patch/update. The digital twin allows for integrity testing of patches and expected changes to the production system. Additionally, system data can be collected from the digital twin to create additional testing scenarios using generative modeling. The System Architecture is shown in Figure \ref{Fig_System_Architecture}.

\subsection{IT devices}
\label{ref_arch_it_devices}
For desktop and laptop computers, Windows is the most commonly used Operating System \cite{DesktopO5:online}. Hence, we use a digital twin for Windows. Our design for uploading a digital twin of an IT device consists of an Image Capture Background Service, a cloud platform where our digital twin is present and a Tests service for the digital twin.

\subsubsection{Image Capture Background Service}
\label{ref_arch_image_capture}
 The background service is responsible for creating and uploading the production system’s image to the cloud. This service keeps on running on the system whose image needs to be captured. The general workflow of this service can be separated into two different functions---that is, the capture and update of device-specific information in the database via the API server and the capture and upload of an image of the system to the cloud storage. These functions are triggered by the end user, typically a sysadmin, on the web application. The web application is the user interface which enables the end users to view their devices, images, digital twins and test reports.

\subsubsection{Digital Twin on Cloud}
\label{ref_arch_digital twin}
 Once the background service uploads an image of the system, it triggers the creation of a virtual machine on the same user interface. This virtual machine once started is our digital twin. The digital twin can be accessed by the user to use as a non-production replica of the production system. On this digital twin, the patch can be applied and used for testing.  
 
\subsubsection{Testing on Digital Twin}
\label{ref_arch_recorder}
Patches can be applied to the digital twin and the system behavior can be observed for sanity testing. This includes observing for any system failure and abnormal behavior. Different tools can be utilized to collect system information both before and after applying patches to observe for any changes. This includes system tools, configuration scanners, and vulnerability scanners to identify conflicts or vulnerabilities that arise. To record manual user testing, this service uses a macro creator tool. However, designing different testing scenarios for the patch can be time-consuming. The data collected by these tools on the digital twin can be used to train generative AI models to generate test scripts for automatic testing, discussed in Section \ref{sec_ai_testing}.

\subsection{IOT devices}
\label{ref_arch__IoT}
The architecture of an IoT based system is different from that of a typical IT system. The Background Service like in IT devices is responsible for creating the image of the IoT device and uploading it to the cloud storage. However, in this case, the image is captured as a tar file. This created image can then be uploaded to the cloud, and can be run as a digital twin on the Docker application. We used Docker as it provides us with a platform to host an operating system initially running on an IoT device. End users can run the image in a Docker container, which is running in a virtual machine on the cloud. Similar to the IT devices, the digital twin for IoT devices can be observed for post-patching system changes and failures. We also collect system data like active processes, cpu and memory usage data from the digital twin of IoT devices to train generative AI models and synthesize additional testing scenarios.

\section{Testing using AI approaches}
\label{sec_ai_testing}
Manually designing different testing scenarios to stress the system require lot of time and effort that are again demanding for smaller businesses with limited resources. The pre-production environment of the digital twins provide an unique opportunity to observe and collect system behavior data. This system data is not only important for sanity testing, but can be used to train Generative AI models. We use the data collected by our digital twins to train Generative  AI Models that can learn the underlying distribution of data and sample a new dataset of synthetic system data. This allows us to design various testing scenarios that can help with stress testing of patches/updates to the system. 

System data is typically captured in two formats: 
\begin{enumerate}
    \item tabular data with a mix of discrete and continuous values
    \item sequence of texts
\end{enumerate}

We demonstrate by training two different generative models for each kind of system data capture. For tabular data, the synthetic data is generated using Generative Adversarial Networks. For sequence of texts, we use the large language model GPT-2. We further discuss these models in this section. 

\subsection{Generative Adversarial Networks}
\label{sec_ai_testing_network}
Generative adversarial Networks (GANs) are a kind of generative model in AI that have been successfully used to generate synthetic data that closely resemble the original data. GANs have been shown to be extremely accurate in synthetic data generation and translation, particularly for image and text data. In case of system data, the data is usually tabular, i.e. a mix of discrete and continuous values. Additionally, the continuous values are not arbitrarily random and usually follow a specific distribution within a given range. To account for this, we need specialized versions of GAN that can accurately replicate system data that is collected over our digital twin. The PriveTab model \cite{kotal2022privetab}, addresses this issue in three key steps (1) Mode-specific normalization, (2) Conditional Generator, and (3) Training by sampling. Additionally, it ensures that the synthetic dataset is distributionally close to the original dataset, such that it can replace the original datasets in functional tests. For this, the Earth Mover's distance (EMD) of the distribution of features in the synthetic is calculated w.r.t. the original dataset. The sampling process continues to sample from the trained generator till the generated distribution is within a threshold distance of the original distribution. We use the model described in PriveTab to generate synthetic system data that are in the tabular format.

\subsection{Generative LLMs}

Generative Pre-trained Transformer 2 (GPT-2) is a deep neural network large language model created by OpenAI and is based on the transformer model. Typically, the architecture of a GPT-2 model is decoder-only transformer \cite{gpt2}. It has decoder blocks stacked on one another. In addition to the feed forward neural network, each decoder stack has a masked self attention, which means that the tokens to the right of the current position or the attention head are not taken into consideration. We used GPT-2 model and fine-tuned it with the cleaned data that we extracted from the macro creator on the digital twin. Fine-tuning enabled the model to perform the downstream task of predicting and generating subsequent commands. The newly generated commands could serve as a new test script in addition to the other test scripts which were created directly by the macro creator.

\section{Prototype for Digital Twin System}
\label{sec_protoype}
For our experiments, we built digital twins for both IT and IoT devices. In the case of IT devices, we created a digital twin for Windows 10 on the Azure cloud. For IoT devices, we created a digital twin for Ubuntu 22.10 running on a Raspberry Pi 4 8GB Model B of ARM v8 architecture which connects to/controls a variety of sensors, and this digital twin runs on a Docker Application. 

\subsection{Digital Twin for IT Devices}
\label{sec_protoype_background_service}
To create a cloud platform supporting the digital twin, specific device information like OS, processor, memory, and storage is captured and updated in our database via the API server. Users add their devices through the web app and then initiate image creation of their device on the web application, followed by the digital twin creation. The image file that is created and uploaded by the background service is in the form of VHD (Virtual Hard Disk) file format because of the support of utility tools such as Microsoft’s Disk2vhd. This cloud platform enables our digital twin to function as a virtual machine on Azure. A macro creator application known as Pulover’s Macro Creator \cite{pmr} was used for the purpose of creating custom test scripts and to train our AI model to generate test scripts in .ahk format. A patch can be applied to the digital twin and the Tests Service can be initiated. Our Windows Tests Service, includes Windows event logs for Application, Security and System. It also includes Vulmap that can be used to test for vulnerabilities and a configuration scanning tool CIS-CAT Lite that can identify the list of changes in configuration required for enhanced security post-patching. The above tests generate test reports. Process Monitor and the test scripts, both custom and generated, can be executed simultaneously to check for any crashes or failures post-patching and will prompt user notifications for debugging. Our Windows 10 Tests Service automates Vulmap, CIS-CAT Lite, Process Monitor, and Windows event logs capturing. The final test report includes vulnerability scan, configuration scan and Windows event logs in a zip file.

\subsection{Digital Twin for IoT Device}
\label{sec_protoype_IoT_device}
To create a digital twin of a Raspberry Pi 4 Model B (ARM v8) running Ubuntu 22.10, we used Docker application and created an image of the IoT device in the form of a tar file instead of a VHD. We have created an Image Capture
Background Service for image creation and uploading. The function of our background service is to create a tar (tar.gz) file of the complete filesystem. Some directories were not included while creating the tar.gz file. For an instance, tmp directory which contains temporary files required by the machine was removed while creating the tar.gz file. A new tmp directory was created while building the final Docker image, from which the Docker container can be started. The created image tar file is also uploaded to the Azure cloud storage by the background service using Azure Command-Line Interface. The virtual
machine is a generic VM and not created from a specialized OS disk like in IT devices. The functionality for creating a generic virtual machine for Ubuntu 22.10 was added to our existing client web application for IT devices. After the image file has been downloaded into the VM, it is imported into the Docker application present in the Azure VM. A new and updated Docker image is built using this imported image. The digital twin was made accessible on the Docker container. The digital twin can be used to perform manual tests to check it’s
functioning. A vulnerability scan can be done using Vulmap for Linux. Other Linux logs for an instance, dmesg, syslog or auth.log can also be monitored to check some of the events that occur. Also, the synthetic active processes and CPU usage data will give us an idea how the new data should look like after the patch is applied to the digital twin.

\begin{table}[]
\centering
\renewcommand{\arraystretch}{1.5}
\resizebox{0.9\columnwidth}{!}{
\begin{tabular}{|l|l|l|l|l|l|}
\hline
\textbf{Process ID} & \textbf{User} & \textbf{\begin{tabular}[c]{@{}l@{}}CPU Usage \\ (in \%)\end{tabular}} & \textbf{\begin{tabular}[c]{@{}l@{}}MEM Usage \\ (in \%)\end{tabular}} & \textbf{\begin{tabular}[c]{@{}l@{}}Time Lapsed\\ (in ms)\end{tabular}} & \textbf{Command} \\ \hline
392                 & root          & 3.3                                                                   & 0.2                                                                   & 00:01.1                                                                & gedit            \\ \hline
1281                & root          & 2                                                                     & 0.1                                                                   & 00:39.5                                                                & top              \\ \hline
184                 & root          & 2                                                                     & 0.1                                                                   & 00:09.1                                                                & sh               \\ \hline
421                 & root          & 1.7                                                                   & 0.7                                                                   & 00:00.7                                                                & xdg-desktop-por  \\ \hline
75                  & root          & 0.7                                                                   & 0.4                                                                   & 00:00.1                                                                & featherpad       \\ \hline
\end{tabular}
}
\caption{Samples of Generated Active Process and CPU Usage data}
\label{tab_synthetic_cpu_usage}
\end{table}

\begin{table}[]
\centering
\renewcommand{\arraystretch}{1}
\resizebox{0.8\columnwidth}{!}{
\begin{tabular}{l|l|l|}
\cline{2-3}
                                                                                                                & \textbf{Recorded Data} & \textbf{Synthetic Data} \\ \hline
\multicolumn{1}{|l|}{\textbf{\begin{tabular}[c]{@{}l@{}}Average CPU Usage \\ (in \%)\end{tabular}}}             & 0.20                   & 0.14                    \\ \hline
\multicolumn{1}{|l|}{\textbf{\begin{tabular}[c]{@{}l@{}}Average Memory Usage \\ (in \%)\end{tabular}}}          & 0.99                   & 1.00                    \\ \hline
\multicolumn{1}{|l|}{\textbf{\begin{tabular}[c]{@{}l@{}}Average Run time \\ of Processes (in ms)\end{tabular}}} & 0.005                  & 0.007                   \\ \hline
\multicolumn{1}{|l|}{\textbf{Most Frequent User}}                                                               & root                   & root                    \\ \hline
\end{tabular}}
\caption{Comparison of Recorded and Generated Active Process and CPU Usage data}
\label{tab_compare_cpu_usage}
\end{table}

\begin{figure*}
        \begin{subfigure}[b]{0.49\textwidth}
           \includegraphics[scale=0.44]{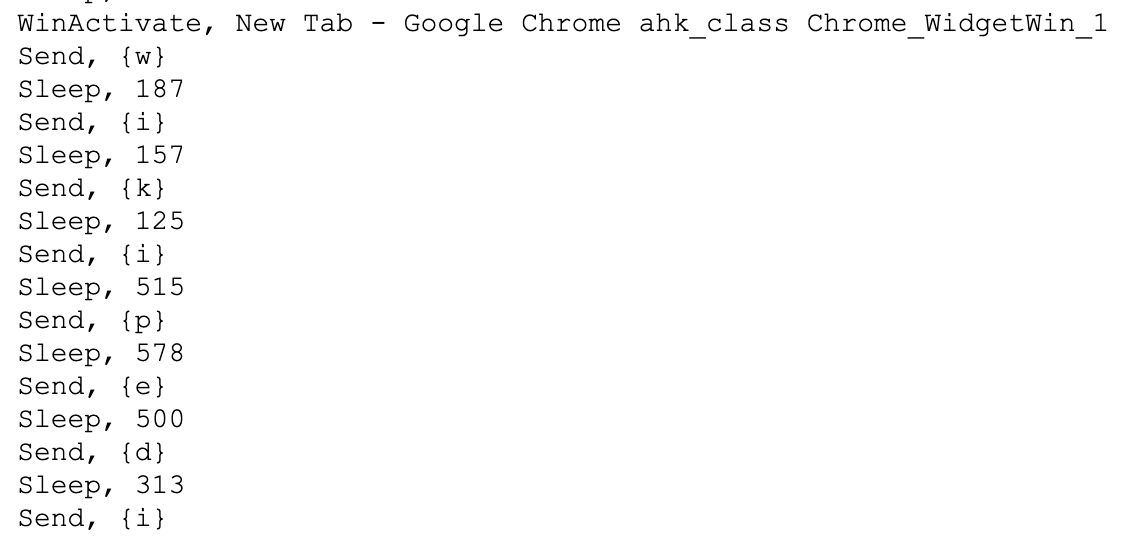}
        \caption{Observed User Interaction Commands}
        \label{fig_realcommand}
        \end{subfigure}
        \begin{subfigure}[b]{0.49\textwidth}
            \includegraphics[scale=0.44]{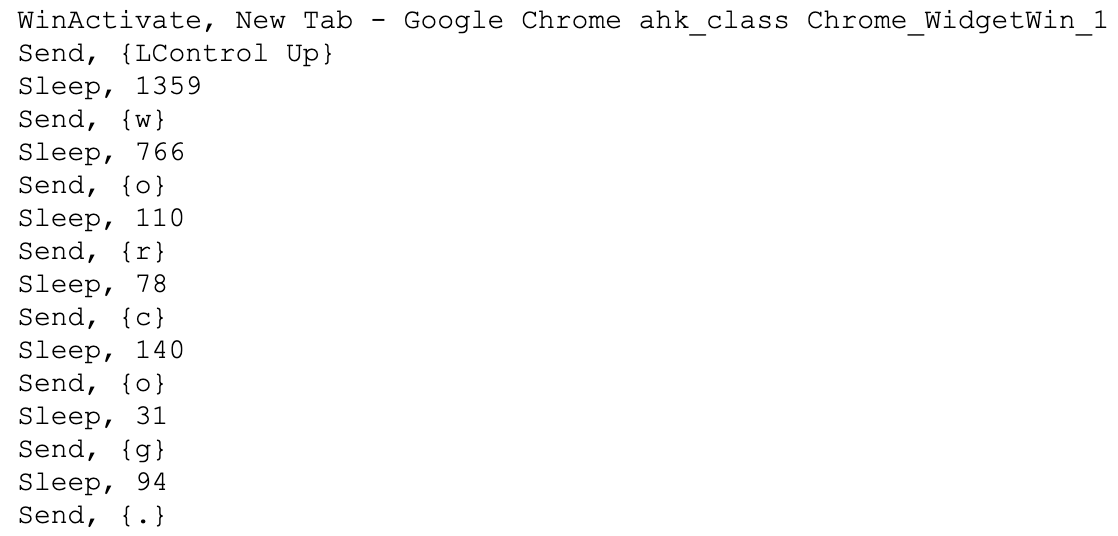}
        \caption{Generated User Interaction Commands}
        \label{fig_syntheticcommand}
        \end{subfigure}
        \caption{Comparison of Observed vs Generated User Interaction Commands}
\end{figure*}

\section{Testing and Validation}
\label{sec_testing}
We collect system data from the digital twins by manually recreating testing scenarios for integrity check of the system. The data collected is used to train generative AI models. In a deployed system, the idea would be to capture system state for a short time as users interact with the system before the patch is applied. System data are typically captured as tabular data or sequence of texts. We demonstrate by collecting CPU usage data in the tabular format and User Interaction Data in the textual format. We demonstrate how the collected data can be used to train generative AI models and sample new sets of data in both formats. In this section, we describe how the data is collected from the digital twin, our results from training the generative models and validate the synthetic dataset against actual data collected from the system.

\subsection{Data Collection}
\subsubsection{Active Processes and CPU usage data}:
As an example of the kind of system data that can be used to determine if the system is behaving normally, we consider process data. The top command is used to get a list of active processes in a Linux system. It provides a dynamic real-time view of processes and threads currently active on Linux kernel. It also includes information about the CPU and memory usage of each active process or thread. This is essential for monitoring system health overview during a patch deployment. The collected data is in a tabular format, where each row represents an active process. The tabular data contains a mix of both continuous and discrete variables. Hence, we use this information to train a GAN model as described in Section \ref{sec_ai_testing_network} and synthesize additional testing scenarios.

\subsubsection{User Interaction Data}:
We have captured user interactions with the digital twin of IT devices. Each set of captured user interactions is a test case scenario, where each set is in the form of a .ahk\cite{ahk} file. We used a macro creator tool to achieve this. The user, first begins the recording, then, for all the applications they want to test, they launch each application and perform various actions using that application. The macro creator then generates a report containing the user's keystrokes, mouse activity, and the timing between actions. The user interactions for each application have been considered as a single test script. We collect the data for an array of such testing scenarios as well. The data collected is cleaned and stored as a sequence of texts.

\subsection{Data Generation} 
\subsubsection{Active Processes and CPU usage data}:
For the CPU usage data, we train a GAN on the collected data. We collected the Active Processes and CPU usage data from the digital twin by running the top command and collecting 560 datapoints. From the collected data, we train the GAN to generate a synthetic dataset of 70 datapoints. Samples from our synthetic CPU usage dataset are provided in Table \ref{tab_synthetic_cpu_usage}. The table consists of features like Process ID, User, CPU Usage, MEM Usage, Time Lapsed and Command to initiate a process. We provide a comparison of the cpu and memory usage from the original vs the generated data in Table \ref{tab_synthetic_cpu_usage}. The average percentage usage of CPU usage per process in the generated table is 0.14 which is close to the observed data, 0.20. The average percentage usage of Memory usage per process in the generated table is 1.00 which is close to the observed average percentage usage of Memory usage per process i.e. 0.99. Also, the average running time of an active process in the synthetic dataset is 0.007 ms which is close to the actual observed average of 0.005 ms and the most frequent user in both datasets were ``root".

\subsubsection{User Interaction Data}
Using Macro Creator tool on the Digital Twin, we capture the User Interaction Data for 6 test scenarios.  Here, each testing scenario is a set of user commands. For an instance, a testing scenario can be on activating Google Chrome and then performing some activities in it. The corresponding user activities  user's keystrokes, mouse activity, and the timing between actions are captured to make one set of testing scenario. We have collected 6 such testing scenarios for training. Using the training dataset, we have fine-tuned a pre-trained GPT-2 model. Given an initial command for a particular test case scenario, the ‘text-generation’ pipeline is able to generate the subsequent commands that are most likely a part of that test case scenario. The generated commands are then converted to a .ahk file and replayed. We generate 3 sets of commands using GPT-2. A sample set of generated commands are shown in Figure \ref{fig_syntheticcommand}. The generated command sequence corresponds to a user typing something on the browser. We compared the generated sequence of text with original sequence of text corresponding to the same scenario, user opening and typing something on the browser. An instance of the original sequence of commands is also provided in Figure \ref{fig_realcommand}. The vectorized generated data had an average cosine similarity score of \textbf{0.73} with the original text sequence. The average Bleu score of the generated text when compared to the original was \textbf{0.72}. 2 of 3 generated testing scenarios using GPT-2 were replayed successfully using AutoHotKey.

\section{Conclusion}
\label{sec_conclusion}
Cybersecurity threats are dynamic and constantly evolving. Fixing vulnerabilities in existing systems and rolling out constant updates and patches are our best defenses against evolving cybersecurity threats. Regular changes to the system require rigorous testing in production and pre-production environments that are often difficult for small and medium-sized businesses who have limited resources. Our framework for creating ``digital twins”, especially for a mix of IT and IoT environments can be used to stage a non-production environment. This can be used for patch management and testing even with limited resources. Furthermore, we use the data collected in two different formats from our digital twins to train generative models. This reduces the need for manual interference during functional testing. In ongoing work, we are transitioning our GPT-2 AI model to GPT-3 or 4, a newer version which has the ability to handle specific topics. Domain specific knowledge graphs have been shown to provide context awareness for ML models \cite{elluri2021policy}, specifically in generation tasks. We are working on adding knowledge guidance into the generative process directly to improve the model performance. 

\section{Acknowledgement}
\label{sec_acknowledgement}
This work was supported in part by MIPS and CyDeploy Inc. We acknowledge the technical feedback from the CyDeploy team in the design of our system. We would also like to acknowledge the efforts of Deviprasad Mohapatra, who was a member of the team that built the digital twin for IT devices described in sections \ref{ref_arch_it_devices} and \ref{sec_protoype_background_service}.

\printbibliography
 \end{document}